# Fractal and Mathematical Morphology in Intricate Comparison between Tertiary Protein Structures

*Ranjeet Kumar Rout[1], Pabitra Pal Choudhury[1], B. S. Daya Sagar[2], Sk. Sarif Hassan[3]*

*[1]Applied Statistics Unit, Indian Statistical Institute, Kolkata, India,*

*[2]System Science and Informatics Unit, Indian Statistical Institute, Bangalore, India.*

*[3]Institute of Mathematics and Application, Andharua, Bhubaneswar-751003, India,*

*Email: ranjeetkumarrout@gmail.com, pabitrapalchoudhury@gmail.com , bsdasagar@isibang.ac.in , sarimif@gmail.com*

## *Abstract*

Intricate comparison between two given tertiary structures of proteins is as important as the comparison of their functions. Several algorithms have been devised to compute the similarity and dissimilarity among protein structures. But, these algorithms compare protein structures by structural alignment of the protein backbones which are usually unable to determine precise differences. In this paper, an attempt has been made to compute the similarities and dissimilarities among 3D protein structures using the fundamental mathematical morphology operations and fractal geometry which can resolve the problem of real differences. In doing so, two techniques are being used here in determining the superficial structural (global similarity) and local similarity in atomic level of the protein molecules. This intricate structural difference would provide insight to Biologists to understand the protein structures and their functions more precisely.

## 1. Introduction:

Proteins are made of amino acids chain with its length ranging from 50 to more than 3000. A carbon atom $C_\alpha$ is connected to a carboxyl (-COOH) group, an amine (-NH2) group, a hydrogen atom and a residue (which depends on the specific amino acid) to formulate a single amino acid. The amine group of an amino acid is covalently bonded by polypeptide bond with the carboxyl group of another amino acid to form a protein. The sequence of $C_\alpha$ carbon atoms forms the backbone of the protein. Whenever the protein is left in its natural environment, it folds to a specific 3D structure. This is due to the forces between the amino acids such that the total free energy is minimized [1]. This renders a stable 3D protein structure. Thus, a protein can either be considered as polypeptides sequence of 20 amino acids occurring naturally or as a tertiary structure into which a particular protein folds [2]. The two highly similar amino acid sequences (primary structures) would not necessarily mean protein functional similarity [2] [3]. Therefore it is evident that comparison of protein functionality requires the intricate comparison of tertiary structures. The search for an effective solution for tertiary protein structural similarity is justified because such tools can be of aid to scientists for prediction of the functions of a newly found protein, in development of procedures for drug design, in the identification of new types of protein architecture, in the organization of the known database of protein structures by classifying them according to their structures and can help to discover unexpected evolutionary and functional inter-relations between proteins [4] [5]. Several algorithms have been devised to compute the similarity between protein structures but they land up in a difficult computational problem as well as accuracy problem [6]. As, in many cases there is not even a single superposition that reveals all regions of similarity between compared proteins (RMSD, DALI, ProSup) [7]. Also, there are many conceptual difficulties associated with various methods (RMSD, ad hoc scores based on local secondary structure, hydrogen bonding pattern, burial status, or interaction environment) which have not been resolved [8]. Classical criteria such as the Root Mean Square Deviation (RMSD) fail to identify similar shapes in a consistent way [9]. To add on various systems have been proposed for structural classification, such as Structural Classification of Proteins (SCOP), Class Architecture Topology Homology (CATH), Families of Structurally Similar Proteins (FSSP), and others. The similarity in their cases is computed using structural alignment algorithms such as DALI, CE, VAST, SSAP and others. Most of these methods are computationally intensive and time-consuming, especially when searching large databases due to intrinsic complexity of structural alignment [10]. Also, the prevailing practice in the protein crystallographic community for computing structural differences is highly inappropriate, in particular when medium- and low-resolution structures are involved [11]. Geometrical feature like Fractal dimension of $C_\alpha$ of the backbone structure of one peptide chain proteins are considered in [12]. Obviously, a more objective method is highly desirable. In this paper these problems have been tried to resolve by introducing two different methods using Mathematical Morphology and Fractals which would yield desired output.

The organization of the paper is as follows: in section 2, the basic review of Mathematical Morphology operations is presented; in section 3, the result and analysis based on the two methods are discussed; conclusion has been made in section 4.

## 2. Basics of Mathematical Morphology and Fractal Dimension.

Mathematical Morphology is a widely used paradigm in the field of image processing. Morphological tools are already very popular for image segmentation, image decomposition etc. Morphological operations like erosion, dilation, opening, closing are used for processing images frequently and produce results with high accuracy. The definitions of these basic morphological operators are as follows [13].

$$\text{Erotion:} \quad A \ominus S = \{a - s : a \in A, s \in S\} = \bigcap_{s \in S} M_s$$

$$\text{Dilation:} \quad A \oplus S = \{a + s : a \in A, s \in S\} = \bigcup_{s \in S} M_s$$

$$\text{Opening}: \quad A \circ S = (A \ominus S) \oplus S$$

$$\text{Closing}: \quad A \bullet S = (A \oplus S) \ominus S$$

Where A denotes the shape that is to be transformed and S denotes the structuring element that is used for the transformation.

### 2.1. Morphological Skeleton.

Morphological skeleton of every geometrical structure is a subset of the original structure which has the same connectivity as the original structure from which inference can be drawn. From each point of the skeleton the distance to the boundary of the original set is the radius of a maximal circle (whose center is at a point of the skeleton) which touches the boundary at least two different points. The skeleton of an object gives a clear idea about the shape of the object. For the shape A, and the structuring element S, the skeleton can be constructed through the operation [14] [15]:

$$Sk_n = (A \ominus nS) \setminus (A \ominus nS) \circ S, \quad \text{for } n = 1, 2, \ldots, N$$

And the reverse process is as follows, Where, N is the number of performed iterations. Dilating the skeleton N times iteratively using the multi-scale structuring elements S a shape that is almost same to the original shape can be achieved.

$$A' = \bigcup_{n=0}^{N} Sk_n \oplus nS$$

Where $nS = S \oplus S \oplus \ldots \oplus S$ (n times).

### 2.2. Fractal Dimension.

A fractal dimension is an index for characterizing fractal patterns or sets. The patterns illustrate self-similarity and the fractal dimension indicates the extent to which the fractal objects fills a particular

Euclidean space in which it is embedded. These dimensions are usually non-integers. The fractal dimensions can be computed through *Box Counting Method* which is briefly stated as follows.

*Box-Counting Method:* This method computes the number of cells required to entirely cover an object, with grids of cells of varying size. Practically, this is performed by superimposing regular grids over an object and by counting the number of occupied cells. The logarithm of N(r), the number of occupied cells, versus the logarithm of $\frac{1}{r}$, where $r$ is the size of one cell, gives a line whose gradient corresponds to the box dimension [16][17]. To calculate the dimension for a fractal *S*, the Box-Counting dimension is defined as,

$$\text{Dim}_{\text{box}}(S) = \lim_{n \to 0} \frac{\log N(r)}{\log \frac{1}{r}}$$

## 3. Methods and Results

In this section two different methods are proposed to compute the similarity between tertiary protein structures in intricate level on the basis of mathematical morphology and fractal dimension.

### 3.0.1. Tertiary Structure Skeleton and its Fractal Dimension.

The Protein Data Bank (PDB) is the largest and most commonly used repository for any kind of information regarding proteins. Information like 3D structure, family, function of every protein found till date is available in PDB. Mainly the X-Ray crystallography and Nuclear Magnetic Resonance is used for determining the 3D structure of the protein. The 3D structure is represented in (x, y, z) coordinates (with respect to an arbitrary origin) of the atoms presented in the protein. The '.pdb' files available in the PDB database contain all the structural information of a protein. Any molecule structure viewer like PyMol, JMol is able to simulate the 3D protein structure available in the .pdb file. The proteins have an intrinsic self-similarity as they are hetero-polymers with a variable composition of twenty different amino acids. Thus, this protein backbone space curve consisting of $C_\alpha$ atoms motivates us to compare 3D protein structure on the basis of their fractal features [16]. Following are the steps for calculating the skeletons and their corresponding fractal dimension.

1. Since the atoms of the tertiary protein structure has three co-ordinates (x, y and z). The idea of slice representation is to decompose a tertiary structure into a sequence of non-overlapping pieces, namely slices, by cutting the tertiary structure on its $m$ vertices with $m'$ planes $p_1, p_2, p_3, \ldots, p_{m'}$ that are perpendicular to the z-axis and the union of planes $m'$ contains all m atoms from the tertiary protein structure. Each slice contains the protein atoms that share the same z-coordinates. Therefore, each slice can be attached with a unique z-interval.

2. Connected component $C_i$ is obtained for all slices by the multi-scale opening of the atoms presented in that slice. $C_i = P_i \circ nS$, $where\ i = 1,\ 2,\ \ldots,\ m$, m is the number of slices and S is the structuring element.

3. Skeleton of the all connected component
$Sk_n(C_i) = (C_i \ominus nS) \backslash (C_i \ominus nS) \circ S$ for $n = 1, 2, \ldots, N$ and $i = 1, 2, 3, \ldots, m$.

4. $A = \cup_{n=0}^{N} Sk_n \oplus nS$   Where, $nS = S \oplus S \oplus \ldots \oplus S$ (n times), where A is the connected component of a tertiary protein structure.

5. Compute the fractal Dimension $D_p$ of the Skeleton $A$ of the corresponding tertiary protein structure.

**Illustration:**

Local similarity of the tertiary structure is obtained through the above steps. In Figure 1, a slice is shown for the protein 2LEP. Each "." represents an atom. The slice contains all the atoms whose z coordinate is within -0.600 to -0.700. After getting a slice, a connected component has been obtained by the multi-scale opening of the atoms presented in that slice. For each iteration of multi-scale opening the size of structuring element increased by one. And for the opening, a primitive structuring element of size $n \times n$ where $n = 1, 2, \ldots N$ is used. By doing this, the tertiary structural atom information has been transformed into two dimensional planes without losing any information at all. The iterations for the slice shown in Figure 1 are shown below.

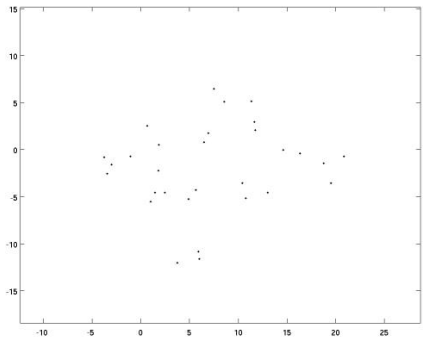

Figure 1  slice for protein 2LEP for $z = -0.600$ to $-0.700$

In Figure 2, $a$ to $j$ shows the multi-scale opening using primitive structuring elements. Starting from size one, each figure shows the iteration with structuring element larger by ten units from the previous one. The iterative opening may take a large number of iterations to contain all the atoms in a particular slice. And there may be more than one plane for a slice. So we dilate the plane with a primitive structuring element. This reduces the number of planes for each slice. The example of the plane after dilating with a disk shape structuring element of size 25, the resulting plane becomes a single connected component as shown in *k* of Figure 2. After acquiring all the planes for a particular protein structure, our next aim is to find the skeletons for each of the plane shapes. Skeleton of the plane- *k* is shown in *l* of Figure 2.

If we stack the skeletons for all the planes over each other, then the resulting image gives us an idea of how the atoms form the overall protein structure in terms of the planes, which are formed by the coordinates of the atoms. This stacked skeleton is basically a projection of the skeleton of the tertiary structure of the protein. For the protein *2LEP* the skeleton structure is shown Figure 3 as given below, From the skeleton we have an idea of fractal-like distribution of protein atoms of the 3D protein structure in the form of plane. Now we can compute the fractal dimension $D_p$ of the skeleton of the corresponding tertiary protein structure and use it as the feature of tertiary protein structure. Fractal Dimension $D_p$ of a group of protein molecules are given in *Table 1* irrespective of their residue.

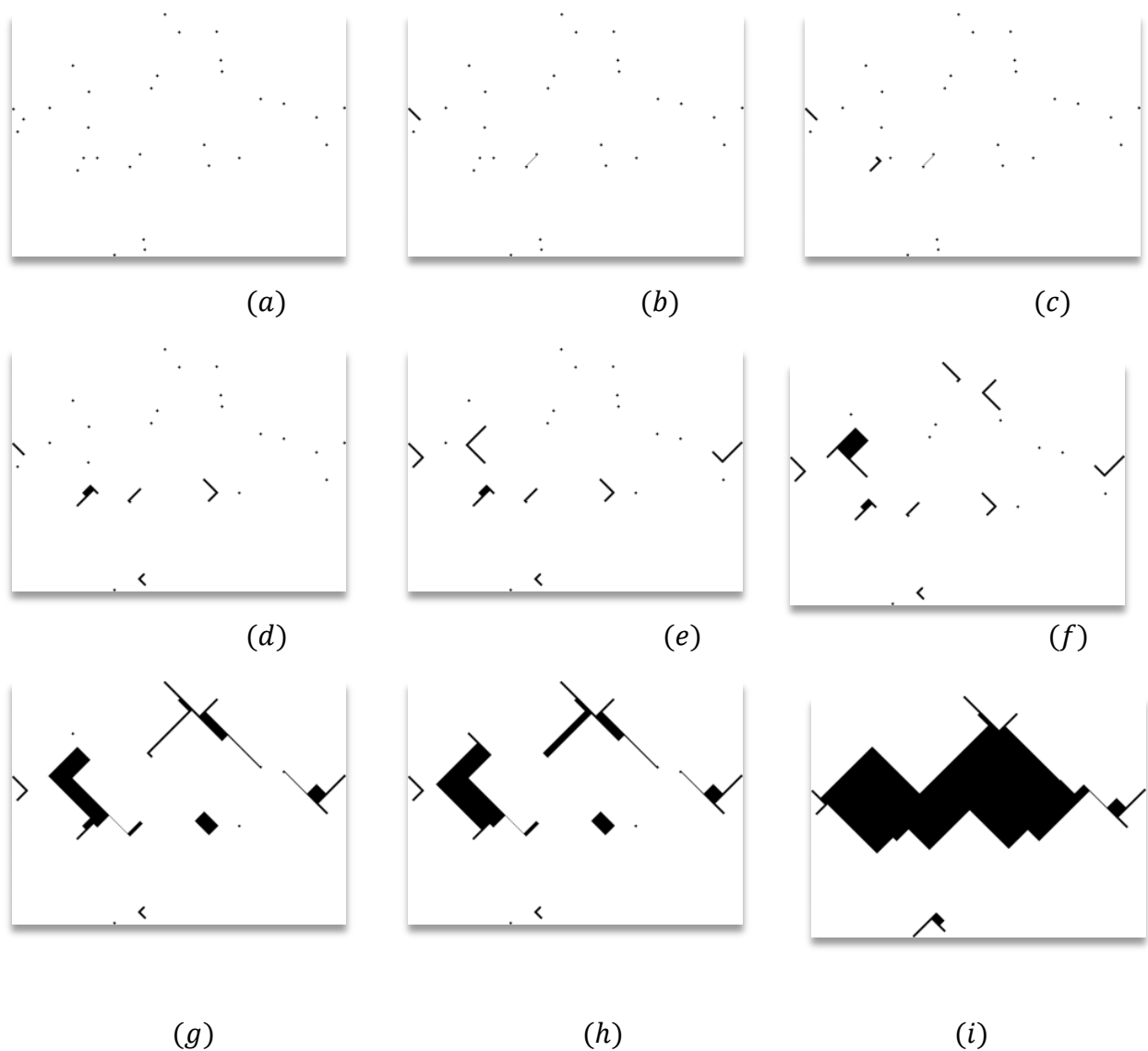

(a)         (b)         (c)

(d)         (e)         (f)

(g)         (h)         (i)

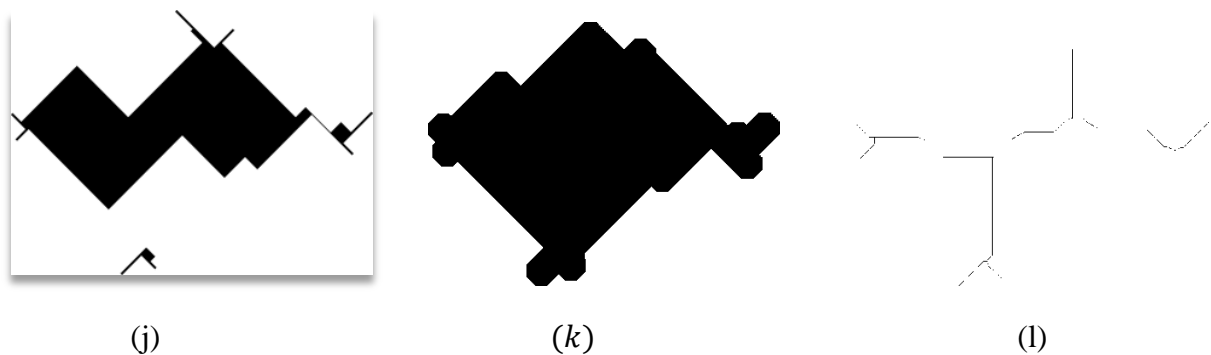

(j)            (k)            (l)

Figure 2. Representation of different iteration of multi-scale opening on One Slice of 2LEP protein (*a* to *k*) and l is the Skeleton of one Slice of 2LEP protein.

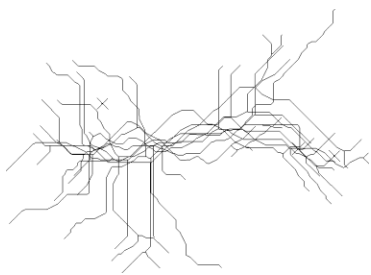

Figure 3: The Stacked Skeleton of 2LEP protein.

**Table 1: Fractal Dimension of Protein Molecules.**

| Protein ID | Residue Number | $D_p$ |
|---|---|---|
| 3v2j | 260 | 1.661190e+000 |
| 3smk | 236 | 1.620140e+000 |
| 3t0o | 238 | 1.646469e+000 |
| 4ecs | 435 | 1.649489e+000 |
| 3v2m | 260 | 1.656160e+000 |
| 3sy1 | 190 | 1.605381e+000 |
| 4ag2 | 226 | 1.695859e+000 |
| 1cah | 259 | 1.661085e+000 |
| 1cai | 259 | 1.660481e+000 |
| 4bij | 476 | 1.680213e+000 |
| 2lep | 69 | 1.549605e+000 |
| 1cgi | 245 | 1.635992e+000 |
| 4eym | 371 | 1.649456e+000 |
| 2cbc | 260 | 1.661399e+000 |

### 3.0.2. Geodesic Dilation and Its Quantification.

From the PDB database the protein structures are viewed by using JMol and the protein structures are rotated depending on the 3-axis, from which we have collected the 6- faces or views (front, left, right, top, bottom, and back) of each 3D protein structure respectively. To find out self similarity between two 2D images we use geodesic dilation which is a morphological transformation to operate only some part of the image (as marker) to grow until the boundary of the image and the advantages of this transformation is that the structuring element can grow at each pixel, according to the image.

The **Geodesic Dilation** $\delta_X$ of an image Y inside X is defined as the intersection of the dilation of Y (with respect to a structuring element S) with the image X

$$\delta^n{}_X = (Y \oplus nS) \cap X \text{ where } n = 1, 2, \ldots, N$$

So Geodesic dilation terminates when all the connected components of X are constructed i.e. idem potency is reached $\forall n > n_0$, $\delta^{(n)}{}_X(Y) = \delta^{(n_0)}{}_X(Y)$. The following step are used for calculating Geodesic Dilation $\delta_p$ of the corresponding faces of protein molecules.

Let $f_1, f_2, \ldots, f_6$ and $g_1, g_2, \ldots, g_6$ are six 2D faces of two 3D protein structures $p_s (source\ protein)$ and $p_t (target\ protein)$.

1. The marker $I_i = f_i \cap g_i$, for all $i = 1, 2, \ldots, 6$.
2. $\delta^n{}_{p_s} = (I_i \oplus nS)! = f_i$, for all $n = 1, 2, \ldots, N$, where S is a structuring element.
3. $\delta^n{}_{p_t} = (I_i \oplus nS)! = g_i$, for all $n = 1, 2, \ldots, N$.
4. The number of dilation $\delta_p = \sum_{i=1}^{6} |\delta^n{}_{p_s} - \delta^n{}_{p_t}|$.

Illustration:

Now we take the front view of two different proteins to compute the global similarity between them. For, this purpose we consider the front view of 3V2J, 3V2M and the common structural part of both the protein molecules, which are given in Figure 3. Here we use **Geodesic dilation** as a parameter for determining the structure similarity between those protein molecules. Now we can determine the number of dilation required from the intersection part to both the protein molecules towards constructing the self similar structure, i.e. the number of geodesic dilation from the marker $I = 3V2J \cap 3V2M$ towards 3V2J and 3V2M is four for each. Similarly for other faces are given in **Table 2.** So the number of geodesic dilation $\delta_p = 2$ with respect to each faces, which shows that 3V2J and 3V2M have more structural similar, whereas for 2LE8 and 2LLS have less structural similarity as $\delta_p = 64$. The **Table 2** given below shows the different geodesic dilation of different protein molecules.

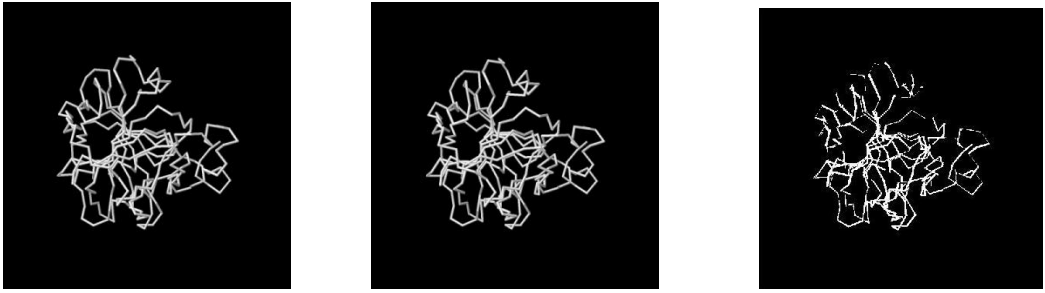

*Front view of 3V2J*      *Front view of 3V2M*     *intersection of 3V2J and 3V2M*

Figure 3: The backbone and intersection part of the protein molecules.

**Table 2:** Geodesic Dilation δ of different faces

| Protein ID | Marker I | Geodesic Dilation δ of different faces | | | | | | $\delta_p$ |
|---|---|---|---|---|---|---|---|---|
| | | Front | Left | Right | Top | Bottom | Back | |
| 3V2J | 3V2J ∩ 3V2M | 4 | 4 | 5 | 6 | 5 | 3 | 2 |
| 3V2M | | 4 | 4 | 4 | 6 | 4 | 3 | |
| 2LE8 | 2LE8 ∩ 2LLS | 10 | 6 | 6 | 7 | 8 | 10 | 64 |
| 2LLS | | 19 | 19 | 21 | 18 | 15 | 19 | |
| 1CAH | 1CAH ∩ 2CBC | 1 | 1 | 1 | 1 | 1 | 1 | 1 |
| 2CBC | | 1 | 1 | 1 | 2 | 1 | 1 | |
| 1CAH | 1CAH ∩ 4EYM | 10 | 10 | 9 | 18 | 18 | 11 | 35 |
| 4EYM | | 20 | 6 | 6 | 9 | 11 | 9 | |
| 3V2J | 3V2J ∩ 3T0O | 12 | 8 | 9 | 12 | 12 | 11 | 35 |
| 3T0O | | 4 | 5 | 5 | 5 | 11 | 5 | |

## 4. Result and Discussion.

In our experiment, we downloaded protein molecules from PDB and the result shows that our methodology performance quite well for comparing tertiary protein structure in intricate level. The similarity between two protein structures *i* and *j* can be computed by using the following equation:

$$\rho = |D_p(i) - D_p(j)|.$$

and Geodesic Dilation $\delta_p$, where $\rho$ the difference between the fractal is dimensions of any two protein molecules and some experimental results are shown in *Table 3*. The difference between the fractal dimensions is essentially measure the difference between the structural complexities. As $\rho$ approaches to zero, the structures closed to be similar. The experimental result shows that if $\rho \leq 0.008$ and $\delta_p \leq 12$, two protein molecules are similar in structures and functions. Thus, lower difference between fractal dimensions and Geodesic dilation will ensure high similarity between the proteins which are being compared. This would become clearer with few examples. For the same from the *Table 3* we conclude that the proteins molecule *1cah* is more similar to *1cai* and *2cbc* as their fractal dimension difference $\rho$ =0.000604, 0.000314 and geodesic dilation $\delta_p = 5\ and\ 1$ respectively.

**Table 3:** Difference between Fractal Dimensions of Compared Protein Pairs

| Protein ID 1 | FD $D_{pi}$ | Protein-ID 2 | FD $D_{pj}$ | ρ | $\delta_p$ | PDB result |
|---|---|---|---|---|---|---|
| **3v2j** | 1.661190e+000 | 3smk | 1.620140e+000 | 0.04105 | 24 | 12% |
| | | 3t0o | 1.646469e+000 | 0.014721 | 35 | 18% |
| | | 4ecs | 1.649489e+000 | 0.011701 | 31 | 2% |
| | | **3v2m** | 1.656160e+000 | **0.00503** | **2** | **100%** |
| | | 3sv1 | 1.605381e+000 | 0.055809 | 28 | 28% |
| | | 4ag2 | 1.695859e+000 | 0.034669 | 39 | 31% |
| **1cah** | **1.661085e+000** | **1cai** | 1.660481e+000 | **0.000604** | **5** | **100%** |
| | | 4bij | 1.680213e+000 | 0.019128 | 21 | 41% |
| | | 2lep | 1.549605e+000 | 0.11148 | 43 | 57% |
| | | 1cgi | 1.635992e+000 | 0.025093 | 35 | 50% |
| | | 4eym | 1.649456e+000 | 0.011629 | 23 | 39% |
| | | **2cbc** | 1.661399e+000 | **0.000314** | **1** | **100%** |

### 4. Conclusion:

In this work, we presented a novel technique to compute the structural similarity of 3D protein structure using fractal dimension and geodesic dilation in atom levels and proteins backbone structure level respectively. Compared with the existing methods, fractal dimension and geodesic dilation is easy to compute and efficient enough to eliminate the limitations encountered in the existing algorithms. In our experiments, atoms of all the protein structures are divided into slices by fixing the *z* co-ordinate value. So only the analysis of the *x-y* planes is done. This work can be further extended by fixing the *x* or y co-ordinate values, i.e. analysis of the *x-z* and *y-z* planes of the protein structure. Thus, this article is allowing us to see the intricate similarity and dissimilarity between tertiary protein structures with enhanced results.